# Integrability and Hamiltonian systems in isotropic turbulence

Zheng Ran[1]

[1]**Shanghai Institute of Applied Mathematics and Mechanics,**
**Shanghai University, Shanghai 200072, P.R.China**

We present developments of the Hamiltonian approach to problems of the freely decay of isotropic turbulence, and also consider specific applications of the modified Prelle-Singer procedure to isotropic turbulence. It demonstrates that a nonlinear second order ordinary differential equation is intimately related to the self-preserving solution of Karman-Howarth equation, admitting time-dependent first integrals and also proving the nonstandard Hamiltonian structure, as well as the Liouville sense of integrability.

The essential characteristic of turbulent flow is that the turbulent fluctuations are random in nature; hence, the final and logical solution of the turbulence problem requires the application of the methods of statistical mechanics [1,2]. To this day, probably the most successful statistical theory of any dynamical problem is the equilibrium statistical mechanics of J.W.Gibbs, L.Boltzmann for classical Hamiltonian systems [3]. It was natural then that the pioneers in turbulence theory would seek some guidance and inspiration from Gibbsian statistical mechanics[4-11]. It was noted that : Turbulence is thus a fundamentally non-equilibrium problem to which the Boltzmann-Gibbs formalism is not directly applicable. New notations and ideas must be introduced. We also noted that : A recent paper by Pradeep et al [12] helped to motivate our work and point to a fruitful line of attack. In a nice paper, Gladwin Pradeep et al present the details of the mathematical backgrounds. By using the modified Prelle-Singer procedure [13], they consider a second order nonlinear ordinary equation, and they identify five new integrable cases. Among these five, four equations admit time-dependent first integrals and remaining one admits time-independent first integral. From the time-independent first integral, nonstandard Hamiltonian structure is deduces, thereby proving the Liouville sense of integrability. Based on this mathematical approach, a Gibbsian statistical mechanics of isotropic turbulence may be developed by taking the scale equation as the Hamiltonian dynamical system analogous to the canonical variables of Hamiltonian systems.

We outline here the details of this approach to demonstrate how to obtain this natural Hamiltonian dynamical system based on the exact solution of the Karman-Howarth equation. We will consider this kind of isotropic turbulence governed by the incompressible Navier-Stokes equations

$$\frac{\partial \vec{u}}{\partial t} + (\vec{u} \cdot \nabla)\vec{u} = -\frac{1}{\rho}\nabla p + \nu \nabla^2 \vec{u} \qquad (1)$$

Where the incompressibility condition reads

$$\nabla \cdot \vec{u} = 0, \qquad (2)$$

The two-point double and triple longitudinal velocity correlation, denoted by $f(r,t)$ and $h(r,t)$ respectively, are defined in a standard way [14]. The Karman-Howarth equation derived from it under the conditions of homogeneity and isotropy reads

$$\frac{\partial}{\partial t}(bf) + 2b^{\frac{3}{2}}\left(\frac{\partial h}{\partial r} + \frac{4h}{r}\right) = 2\nu b\left(\frac{\partial^2 f}{\partial r^2} + \frac{4}{r}\frac{\partial f}{\partial r}\right) \tag{3}$$

where $(r,t)$ is the spatial and time coordinates, $\nu$ is the kinematics viscosity, and $b = <u^2>$ denotes the turbulence intensity. Following von Karman and Howarth, we introduce the new variables

$$\xi = \frac{r}{l(t)} \tag{4}$$

Where $l = l(t)$ is a uniquely specified similarity length scale. As already noted in the freely decay isotropic turbulence case [15-19], equation (3) could be reduced into the following systems:

[1] Two-point double and triple longitudinal velocity correlation, denoted by $f(r,t)$ may be written

$$\frac{d^2 f}{d\xi^2} + \left(\frac{4}{\xi} + \frac{a_1}{2}\xi\right)\frac{df}{d\xi} + \frac{a_2}{2}f = 0 \tag{5}$$

With boundary conditions $f(0) = 1$, $f(\infty) = 0$, and $a_1$ and $a_2$ are constant coefficients.

[2] The turbulent scales equations read

$$\frac{dl}{dt} = a_1 \cdot \frac{\nu}{l} + 2I_1 \cdot \sqrt{b} \tag{6}$$

$$\frac{db}{dt} = -a_2 \cdot \frac{\nu b}{l^2} \tag{7}$$

Where $I_1$ is a constant of integration.

[3] For third order of correlation coefficient, from the system of Karman-Howarth equations one can derive the following equation

$$\frac{dh}{d\xi} + \frac{4}{\xi}h = -\frac{l}{2b^{\frac{3}{2}}}\frac{db}{dt}f + \frac{1}{2\sqrt{b}}\frac{dl}{dt}\xi\frac{df}{d\xi} + \frac{\nu}{\sqrt{b}l}\left(\frac{d^2 f}{d\xi^2} + \frac{4}{\xi}\frac{df}{d\xi}\right) \tag{8}$$

In fact, the problem is thus reduce to solving a soluble system, furthermore, one can prove that there is a self-closed second order nonlinear dynamical system for $l = l(t)$:

$$\frac{d^2l}{dt^2} = -(2a_1 + a_2) \cdot \frac{v}{2l^2} \cdot \frac{dl}{dt} + (a_1 a_2 v^2) \cdot \frac{1}{2l^3} \tag{9}$$

Let

$$z = l^{-2} \tag{10}$$

Or

$$l = z^{-\frac{1}{2}} \tag{11}$$

The scale equation for $z = z(t)$ is:

$$\frac{d^2z}{dt^2} - \left\{\frac{3}{2}\right\} z^{-1} \cdot \left[\frac{dz}{dt}\right]^2 + (2a_1 + a_2) \cdot \frac{v}{2} \cdot z \cdot \left\{\frac{dz}{dt}\right\} + (a_1 a_2 v^2) \cdot z^3 = 0. \tag{12}$$

The most important finds is that :By using the modified Prelle-Singer procedure, we identify the new integrable cases in the above equation. The turbulence scale equations admit time dependent first integrals and the remaining one admits time-independent first integral. From the time-independent first integral, nonstandard Hamiltonian structure is deduced, thereby proving the Liouville sense of integrability. In the case of time-dependent integrals, we either explicitly integrate the system or transform to a time-independent case and deduce the underlying Hamiltonian structure.

Let $T$ and $L$ being respectively a characteristic time scale and length scale. The Reynolds numbers used to characterize the flow are

$$R_e = \frac{UL}{v} = \frac{L^2}{vT}. \tag{13}$$

The only parameter that enters the equations is the kinematics viscosity $v$. The resulting equation can be written .The time and length in no dimensional form are $\tau, l_1$. Let

$$A = \frac{2a_1 + a_2}{2} \cdot \frac{1}{R_e}, \tag{14}$$

$$B = \frac{1}{2} \cdot a_1 a_2 \cdot \left[\frac{1}{R_e}\right]^2. \tag{15}$$

To further consideration the following variable is introduced:

$$A = \left(1 + \frac{a_2}{2a_1}\right) \cdot \frac{a_1}{R_e}, \tag{16}$$
$$\equiv \mu(1 + \sigma)$$

$$B = \frac{a_2}{2a_1} \cdot \left[\frac{a_1}{R_e}\right]^2. \tag{17}$$
$$\equiv \sigma\mu^2$$

To further consideration

$$w = l_1^{-2}. \tag{18}$$

Consequently the equation (12) can be rewritten

$$\frac{d^2 w}{d\tau^2} + \frac{k_1}{w}\left[\frac{dw}{d\tau}\right]^2 + k_3 w \cdot \left[\frac{dw}{d\tau}\right] + k_4 w^3 = 0. \tag{19}$$

Where

$$k_1 = -\frac{3}{2}, \tag{20}$$

$$k_3 = A, \tag{21}$$

$$k_4 = 2B. \tag{22}$$

By using the modified Prelle-Singer procedure, we identify the new integrable cases in this equation. From the time independent first integral, nonstatandard Hamiltonian structure is deduced thereby proving the Liouville sense of integrability.

The parameter choice condition read as

$$\begin{aligned} \Delta &\equiv 4k_4(2+k_1) - k_3^2 \\ &= -\frac{(2a_1 - a_2)^2}{4} \cdot R_e^{-2} . \\ &\leq 0 \end{aligned} \tag{23}$$

We use the canonical transformation

$$w = \frac{U}{P}, \tag{24}$$

$$p = \frac{1}{2}P^2. \tag{25}$$

Where $p$ is the canonical momentum defined by

$$p = w^{(2-r)k_1} \cdot \left\{\frac{dw}{d\tau} + \frac{k_3(r-1)w^2}{r(2+k_1)}\right\}^{1-r}. \tag{26}$$

For the choice $\Delta < 0$, the Hamiltonian can be written in terms of the new canonical variables as

$$H = 2(2+k_1)rr_{12}\left[P^{2+k_1}U^{-k_1}\right]^{r_{12}} - 2^{r_{12}}k_3(r-1)U^2. \tag{27}$$

The associated canonical equations of motion now become

$$\frac{dU}{d\tau} = \frac{2(2+k_1)^2 rr_{12}^2}{P}\left[\frac{P^{2+k_1}}{U^{k_1}}\right]^{r_{12}}, \tag{28}$$

$$\frac{dP}{d\tau} = 2^{r_{12}}k_3(r-1)2U + \frac{2k_1(2+k_1)rr_{12}^2}{U}\left[\frac{P^{2+k_1}}{U^{k_1}}\right]^{r_{12}}. \tag{29}$$

Where

$$r = \frac{k_3^2 \pm k_3\sqrt{k_3^2 - 4k_4(2+k_1)}}{2k_4(2+k_1)}, \tag{30}$$

$$r_{12} = \frac{r-1}{r-2}. \tag{31}$$

Rewriting (28) for

$$P = \left[\frac{U^{k_1 r_{12}} U'}{2(2+k_1)^2 r r_{12}^2}\right]^{\frac{1}{(2+k_1)r_{12}-1}}, \tag{32}$$

And substituting the latter into (27) we get

$$H = \mu_2 U^{m_1}[U']^{m_2} + \mu_3 U^2 \equiv E. \tag{33}$$

Where

$$\mu_2 = \frac{2(2+k_1)r r_{12}}{\left[2(2+k_1)^2 r r_{12}^2\right]^{m_3(2+k_1)r_{12}}}, \tag{34}$$

$$\mu_3 = -2^{r_{12}} k_3 (r-1), \tag{35}$$

$$m_1 = k_1[(2+k_1)m_3 - r_{12}], \tag{36}$$

$$m_2 = r_{12}(2+k_1)m_3. \tag{37}$$

This in turn can be brought to the form

$$\frac{dU}{d\tau} = \left\{\frac{E - \mu_3 U^2}{\mu_2 U^{m_1}}\right\}^{\frac{1}{m_1}}. \tag{38}$$

Now integrating the above equation we get

$$\tau - \tau_0 = \frac{m_2 U}{m_1 + m_2}\left[\frac{\mu_2 U^{m_1}}{E}\right]^{\frac{1}{m_2}} F\left[\frac{m_1 + m_2}{2m_2}, \frac{1}{m_2}, \frac{m_1 + 3m_2}{2m_2}, \frac{\mu_3 U^2}{E}\right]. \tag{39}$$

Where $F[a,b,c,z]$ is the hyper geometric function and $t_0$ is an integration constant.

We use the same canonical transformation, and rewrite the Hamiltonian with the parametric choice $\Delta = 0$ as

$$H = \log\left[\frac{U^{k_1}}{P^{k_1}}\right] - \log[P^2] + \frac{k_3}{4(2+k_1)}U^2. \tag{40}$$

The associated canonical equations of motion now become

$$\frac{dU}{d\tau} = -\frac{2+k_1}{P}, \tag{41}$$

$$\frac{dP}{d\tau} = -\frac{k_1}{U} - \frac{2k_3}{4(2+k_1)}U. \tag{42}$$

Now integrating Eqs.(41), we get

$$\tau - \tau_0 = \tilde{E} U^{\frac{2(1+k_1)}{2+k_1}} \cdot \left[-\frac{k_3 U^2}{(2+k_1)^2}\right]^{-\frac{1+k_1}{2+k_1}} \Gamma\left[\frac{1+k_1}{2+k_1}, -\frac{k_3}{4(2+k_1)^2} U^2\right], \qquad (43)$$

Where

$$\tilde{E} = -2^{\frac{4+4k_1}{2+k_1}} \exp\left[-\frac{E}{2+k_1}\right], \qquad (44)$$

$\Gamma$ Is the gamma function.

One of the main goals in the development of the theory of Hamiltonian dynamical system has been to make progress in understanding the statistical behavior of turbulence. The attempts to relate turbulence to Hamiltonian system motion has received strong impetus from the celebrated papers by Burgers, Hopf , and T.D.Lee and others. Considerable success has been achieved mainly at the wave number space, where we take the real and imaginary parts of the wave vector components of the velocity fields as phase space coordinates analogous to the canonical variables of Hamiltonian systems. For fully developed turbulence, many questions remain unanswered. The aim of this letter was to show that there are Hamiltonian dynamical systems that are much simpler than the former, but that can still have turbulent states and for which many concepts developed in the theory of Hamiltonian system dynamical systems can be successfully applied. In this connection we advocate a broader use of the universal properties of a wide range of isotropic turbulence phenomena .

The work was supported by the National Natural Science Foundation of China (Grant Nos.11172162, 10572083).